\documentclass[tightenlines,aps,prl]{revtex4}
\usepackage{graphicx}
\usepackage{dcolumn}
\usepackage{bm}
\usepackage{epsfig}
\usepackage{slashed}
\newcommand{\be}{\begin{equation}}
\newcommand{\ee}{\end{equation}}
\newcommand{\bea}{\begin{eqnarray}}
\newcommand{\eea}{\end{eqnarray}}
\newcommand{\bean}{\begin{eqnarray*}}
\newcommand{\eean}{\end{eqnarray*}}

\newcommand{\gapproxeq}{\lower
.7ex\hbox{$\;\stackrel{\textstyle >}{\sim}\;$}}
\newcommand{\lapproxeq}{\lower
.7ex\hbox{$\;\stackrel{\textstyle <}{\sim}\;$}}

\begin{document}

\bibliographystyle{unsrt}

\title{ Towards a dynamical understanding of the non-$D \bar D$ decay of $\psi(3770 )$}

\author{Yuan-Jiang Zhang$^{1}$, Gang Li$^1$, Qiang Zhao$^{1,2,3}$}

\affiliation{1) Institute of High Energy Physics, Chinese Academy of
Sciences, Beijing 100049, P.R. China \\
2) Department of Physics, University of Surrey, Guildford, GU2 7XH,
United Kingdom \\
3) Theoretical Physics Center for Science Facilities, CAS, Beijing
100049, China}

\date{\today}

\begin{abstract}

We investigate the $\psi(3770)$ non-$D\bar{D}$ decays into $VP$,
where $V$ and $P$ denote vector and pseudoscalar mesons,
respectively, via OZI-rule-evading intermediate meson rescatterings
in an effective Lagrangian theory. By identifying the leading meson
loop transitions and constraining the model parameters with the
available experimental data for $\psi(3770)\to J/\psi\eta$,
$\phi\eta$ and $\rho\pi$, we succeed in making a quantitative
prediction for all $\psi(3770)\to VP$ with $BR_{VP}$ from $0.41\%$
to $0.64\%$. It indicates that the OZI-rule-evading long-range
interactions are playing a role in $\psi(3770)$ strong decays, and
could be a key towards a full understanding of the mysterious
$\psi(3770)$ non-$D\bar{D}$ decay mechanism.

\end{abstract}

\maketitle


\vspace{1cm}


Charmonium state $\psi(3770)$ has a mass just above the open
$D\bar{D}$ threshold, which allows it to decay into charmed mesons,
i.e. $D\bar{D}$, without the so-called Okubo-Zweig-Iizuka (OZI)
rule~\cite{OZI} suppression. This scenario qualitatively explains
that the width of the $\psi(3770)$ is about two orders of magnitude
larger than those of the $J/\psi$ and $\psi^\prime$ due to the
dominant $D\bar{D}$ decay. An interesting and nontrivial question
here is whether the $\psi(3770)$ decay is totally saturated by
$D\bar{D}$, or whether there exist significant non-$D\bar{D}$ decay
channels. Unfortunately, a definite answer from either experiment or
theory is unavailable. CLEO Collaboration measured the exclusive
cross sections for $\psi(3770)\to D\bar{D}$ \cite{He:2005bs,:2007zt}
and inclusive cross sections for $\psi(3770)\to$
hadrons~\cite{Besson:2005hm}. These results lead to
$BR_{\psi(3770)\to D\bar{D}}=(103.0\pm 1.4 ^{+5.1}_{-6.8})\%$, of
which the lower bound suggests the maximum non-$D\bar{D}$ branching
ratio is about 6.8\%.

The $D\bar{D}$ production cross sections measured by
BES~\cite{bes-3770}
are consistent with CLEO~\cite{:2007zt}. However, the analyses lead
to much larger non-$D\bar{D}$ branching ratios of $\sim 15\%$. Such
a significant discrepancy makes the experimental status quite
puzzling. Also, the search for exclusive non-$D\bar{D}$ decays has
been carried out at both CLEO~\cite{cleo-exclusive}
and BES~\cite{Bai:2003hv}. In Ref.~\cite{pdg2008}, three
non-$D\bar{D}$ hadronic decay branching ratios are listed, i.e.
$\psi(3770)\to J/\psi \pi\pi$, $J/\psi\eta$ and $\phi\eta$, while
tens of other channels have only experimental upper limits due to
the poor statistics. In the radiative decay channel, $\psi(3770)\to
\gamma\chi_{c0}$ and $\gamma\chi_{c1}$ are listed while an upper
limit is given to $\gamma\chi_{c2}$. The sum of those channels,
however, is far from clarifying the mysterious situation of the
$\psi(3770)$ non-$D\bar{D}$ decays. It hence stimulates intensive
experimental and theoretical
efforts~\cite{Kuang:1989ub,Ding:1991vu,Rosner:2001nm,Rosner:2004wy,Eichten:2007qx,Voloshin:2005sd,Achasov:2005qb,He:2008xb}
on understanding the nature of $\psi(3770)$ and its strong and
radiative transition dynamics.

In this Letter we propose that the dominant $D\bar{D}$ decay is
strongly correlated with the non-$D\bar{D}$ ones. We argue that the
intermediate $D\bar{D}$ and $D\bar{D^*}+c.c.$ rescatterings, which
annihilate the $c\bar{c}$ at relatively large distance by the
OZI-rule evading processes, may provide a natural mechanism for
quantifying the $\psi(3770)$ non-$D\bar{D}$ decays.

\begin{figure}
\begin{center}
\epsfig{file=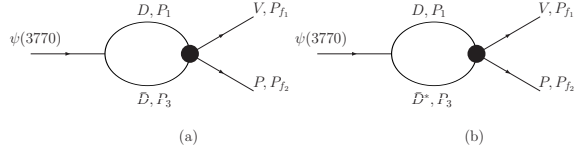, scale=0.4} \caption{Schematic diagrams for
the charmed meson rescatterings into a non-$D\bar{D}$ decay channel
$VP$ via (a) $D\bar{D}$ loop and (b) $D\bar{D}^*$. The conjugation
channel $D^*\bar{D}$ is also implied in (b).} \protect\label{fig-1}
\end{center}
\end{figure}

As illustrated in Fig.~\ref{fig-1} the $c\bar{c}$ pair first couples
to an intermediate meson pair, e.g. $D\bar{D}$, and then these two
mesons rescatter into two light mesons via the $c\bar{c}$
annihilation and a light quark pair creation. Qualitatively, with
the branching ratio for $\psi(3770)\to D\bar{D}$ at an order of one,
the rescattering process could be suppressed by two or three orders
of magnitude. Note that the OZI-evading rescatterings are open to
numerous final-state light mesons. It might be possible that a sum
of those exclusive final states would account for a sizeable
fraction of the $\psi(3770)$ branching ratios.

\begin{figure}
\begin{center}
\epsfig{file=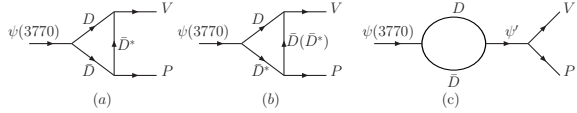, scale=0.4} \caption{The $t$ [(a) and (b)] and
$s$-channel (c) meson loops in $\psi(3770)\to VP $.}
\protect\label{fig-2}
\end{center}
\end{figure}

A natural way of describing the rescattering processes is to expand
the amplitude in Fig.~\ref{fig-1} via the Mandelstam variables
$t\equiv (P_{f1}-p_1)^2$ and $s\equiv
(P_{f1}+P_{f2})^2=M^2_{\psi(3770)}$. At leading order, the
$t$-channel is via an additional meson exchange transition, while
the $s$-channel can be recognized as the vector meson mixings, e.g.
$\psi(2S)$-$\psi(1D)$ mixing~\cite{Ding:1991vu,Rosner:2001nm}. The
typical transition diagrams are shown in Fig.~\ref{fig-2}. The
intermediate $D\bar{D}$ rescattering will contribute to the
absorptive part of the transition amplitude and is not to be dual to
the pQCD leading transition via short-range gluon exchanges. This is
an explicit indication that long-range interactions can play an
important role in such a transition. The intermediate
$D\bar{D^*}+c.c.$ can contribute to the real part of the transition
amplitude due to its large coupling to
$\psi(3770)$~\cite{Cheng:2004ru} and the break-down of the local
quark-hadron duality~\cite{Lipkin:1986bi,Lipkin:1986av}. By
clarifying the above points, we are ready to construct the theory
for probing the role played by the intermediate charmed meson loops
in $\psi(3770)\to VP$.

The following effective Lagrangians are needed in the evaluation of
the $t$ and $s$-channel transitions,
 \bea
\mathcal{L}_{\psi D\bar D} &=& g_{ \psi D\bar D}\{D
\partial_\mu {\bar D}-\partial_\mu D\bar{D} \}{\psi^\mu},\nonumber\\
 \mathcal{L}_{\mathcal V D {\bar D}^\ast} &=& -i
g_{\mathcal V D {\bar D}^\ast} \epsilon_{\alpha\beta\mu\nu}
\partial^\alpha \mathcal{V^\beta} \partial^\mu \bar {D^\ast}^\nu D + H.c. ,\nonumber\\
 \mathcal{L}_{\mathcal P D^\ast {\bar D}^\ast} &=& -i
g_{\mathcal P D^\ast {\bar D}^\ast} \epsilon_{\alpha\beta\mu\nu}
\partial^\alpha D^{\ast\beta} \partial^\mu \bar{D}^{\ast\nu}  \mathcal{P} + H.c. ,
 \nonumber \\
 \mathcal{L}_{\mathcal P \bar D D^\ast} &=& g_{D^\ast \mathcal P
 \bar D}\{\bar D
\partial_\mu {\mathcal P}-\partial_\mu {\bar D} \mathcal P\} D^{\ast\mu} +H.c.,
 \eea
where $\epsilon_{\alpha\beta\mu\nu}$ is the Levi-Civita tensor; $
\mathcal P $ and $ \mathcal V^\beta$ are the pseudoscalar and vector
meson fields, respectively.

The charmed meson couplings to light meson are obtained in the
chiral and heavy quark limits~\cite{Cheng:2004ru},
 \bea
 g_{D^\ast D \pi} = \frac{2}{ f_\pi } g
\sqrt{m_D m_{D^\ast}} , &  &   g_{D^\ast D^\ast \pi} =
\frac{g_{D^\ast D \pi}}{\tilde{M_D}} ,\nonumber\\
g_{D^\ast D \rho} = \sqrt{2} \lambda g_\rho , && g_{D D \rho} =
{g_{D^\ast D \rho}} \tilde{M_D} ,
 \eea
where $f_\pi$ = 132 MeV is the pion decay constant, and
$\tilde{M_D}\equiv \sqrt{m_D m_{D^\ast}}$ sets a mass scale. The
parameters $g_\rho$ respects the relation $g_\rho = {m_\rho /
f_\pi}$~\cite{Casalbuoni:1996pg}. We take $\lambda = 0.56\,
\text{GeV}^{-1} $ and  $g = 0.59$ ~\cite{Liu:2006dq,Yan92}.

The coupling $g_{\psi(3770) D \bar D}$ is extracted by,
 \be \Gamma_{\psi(3770)\to D \bar D} =
\frac{g_{\psi(3770) D \bar D}^2 |\vec{p}\,|^3}{6\pi
M_{\psi(3770)}^2},
 \ee
 where $|\vec{p}\,|$ is the D-meson momentum. The branching ratios
for $\psi(3770) \to D^+ D^-$ and $D^0 \bar D^0$ are slightly
different. They give $g_{\psi(3770) D^+ D^-} =12.71$ and $
g_{\psi(3770) D^0 \bar D^0}=12.43$, and reflects the isospin
violation due to the mass difference between the $u$ and $d$ quark.
Taking into account the consequent kinematic difference, we also
have access to isospin violating channels via the meson loops.

For other couplings, we take the SU(3) flavor symmetry as a leading
order approximation which leads to $g_{{D^0}^\ast \bar D^0 \,u\bar
u} = g_{{D^+}^\ast D^- \,d\bar d} = g_{{D_s^+}^\ast D_s^- \,s\bar
s}$ \,and \,$g_{{D^0}^\ast \bar {D^0}^\ast \,u\bar u} =
g_{{D^+}^\ast {D^-}^\ast \,d\bar d} = g_{{D_s^+}^\ast {D_s^-}^\ast
\,s\bar s}$. So we have  $g_{D^\ast D \pi} = \sqrt{2} g_{D^\ast D
q\bar q(0^-)}, \ g_{D^\ast D \rho} = \sqrt{2} g_{D^\ast D q\bar
q(1^-)}, \ g_{D^\ast D  s\bar s} = 0$, and $g_{{D_s}^\ast D_s n\bar
n} = 0$,
 with
$n$ for $u$ or $d$ quark. Similar relations are also implied for
$g_{D^\ast D^\ast \pi}$, and $g_{D D \rho}$.

We adopt coupling constants $g_{J/\psi D {D^\ast}} = 3.84$
GeV$^{-1}$ and $ g_{J/\psi D {D}} = 7.44 $ from
Ref.~\cite{Oh:2007ej}. Coupling $g_{\psi(3770) D \bar D^\ast }$ can
be related to $g_{\psi(3770) D \bar D }$ via ${g_{\psi(3770) D \bar
D^\ast }} = g_{\psi(3770) D \bar D }/ \tilde{M_D}$.

The $\eta$-$\eta^\prime$ mixing is considered in a standard way,
 \bea
 \eta &=& \cos\alpha_P |n\bar n\rangle  - \sin\alpha_P |s\bar s\rangle,
 \nonumber\\
 \eta^\prime &=& \sin\alpha_P |n\bar n\rangle + \cos\alpha_P |s \bar s\rangle ,
 \eea
 where $|n\bar{n}\rangle \equiv |u\bar{u}
 +d\bar{d}\rangle/\sqrt{2}$, and the mixing angle $\alpha_P=\theta_P +
\arctan(\sqrt{2})$ with $\theta_P\simeq -24.6^\circ$ or $\sim
-11.5^\circ$ for linear or quadratic mass relations,
respectively~\cite{pdg2008}.  We adopt $\theta_P = -19.1^\circ$
~\cite{Liu:2006dq}.

By investigating $\psi(3770) \to J/\psi\eta$, $\phi\eta$ and
$\rho\pi$ simultaneously, we expect to obtain constraints on the
theory by which we can then make predictions for other $VP$
channels. Although these decays are OZI-rule-suppressed processes,
their kinematics are slightly different. The production of $J/\psi$
in $\psi(3770)\to J/\psi\eta$ suggests that it is a very soft
process. The momentum carried by the final state meson in the
$\psi(3770)$-rest frame is $p=0.359$ GeV which is much less than the
masses of both $\eta$ and $J/\psi$. Thus, we argue that
$\psi(3770)\to J/\psi\eta$ is dominated by the intermediate meson
loops. Note that the $t$-channel loops suffer from
divergence~\cite{Zhang:2008ab}. We then introduce a cut-off in the
loop integrals via a standard dipole form factor,
 \be\label{ff-loop}
 {\cal F}(q^2)=\left(\frac{\Lambda^2-m_{ex}^2}{\Lambda^2-q^2}\right)^2 \ ,
 \ee
 where $\Lambda \equiv  m_{ex} +
\alpha \Lambda_{QCD}$, with $\Lambda_{QCD} = 0.22$ GeV;  $m_{ex}$ is
the mass of the exchanged meson and $\alpha$ is a parameter to be
determined by experimental data for $\psi(3770)\to J/\psi\eta$.

The $s$-channel meson loop contributions can be determined via the
on-shell approximation. We find that the branching ratio given by
the $\psi^\prime$-$\psi(3770)$ mixing in $\psi(3770)\to J/\psi\eta$
is $BR=1.3\times 10^{-5}$ which is much smaller than the
$t$-channel, and indicates the dominance of the $t$-channel. With
$BR^{exp}_{J/\psi\eta}=(9.0\pm 4)\times 10^{-4}$~\cite{pdg2008}
$\alpha=1.73$ can be determined and the exclusive $t$-channel
contributes $8.44\times 10^{-4}$.

As follows, we fix $\alpha=1.73$ in the form factors as an overall
parameter. Two aspects must be taken care of here. Firstly, since
relatively large momentum transfers are involved in $\psi(3770)$
decays into light $VP$, the pQCD leading contribution via SOZI
transitions may play a role. This part contributes to the real part
of the transition amplitude and will not be dual with the long-range
intermediate meson loops as recognized by the absorptive feature of
the $D\bar{D}$ rescattering in the on-shell approximation. Secondly,
for those light $VP$ decay channels, their SOZI amplitudes can be
related to each other by  the flavor-blind
assumption~\cite{Seiden:1988rr,Li:2007ky} for quark-gluon coupling,
 \bea
&g_S^{\rho^{0}\pi^{0}}:g_S^{K^{*+}K^{-}}: g_S^{\omega\eta}:
g_S^{\omega\eta'}:g_S^{\phi\eta}:g_S^{\phi\eta'}\nonumber\\
=& 1 :1 : \cos\alpha_P:\sin\alpha_P: (-\sin\alpha_P): \cos\alpha_P \
, \eea with the other isospin channels implied.

The transition amplitude for  $\psi(3770) \to VP$ can be expressed
as
 \bea {\mathcal M}_{fi} &=&
{\mathcal M}^{L}+ e^{i\delta} {\mathcal M}^{SOZI}\equiv i(g_{L}+ e^{i\delta} g_S {\cal F}_S({\vec p}_V)) \nonumber\\
& & \times   \varepsilon_{\alpha\beta\mu\nu} P_\psi^\alpha
\epsilon_\psi^\beta P_V^\mu \epsilon_V^{*\nu}/M_{\psi(3770)}
 \eea
where the property of antisymmetric tensor is applied to factorize
out the effective couplings in the second line and $\delta$ is the
phase angle between the meson loop and SOZI amplitudes. A
conventional form factor, $ {\cal F}_S^2({\vec P}_V) \equiv \exp
({-{\vec P}_V^2/{8\beta^2}})$ with $\beta = 0.5\mbox {GeV}$, is
applied for the SOZI transition with ${\vec P}_V$ the final three
momentum in the $\psi(3770)$ rest frame~\cite{close-et-al,Li:2007ky}

With $\alpha=1.73$ fixed, we can then determine the other two
parameters $g_S\equiv g_S^{\rho^{0}\pi^{0}}=0.085$ and
$\delta=-66^\circ$ by experimental data, i.e. $BR_{\phi\eta}=(3.1\pm
0.7)\times 10^{-4}$~\cite{pdg2008} and $BR_{\rho\pi}<0.24\%$ with
C.L. of 90\%~\cite{Ablikim:2005cd}. In Tab.~\ref{tab-1} theoretical
predictions for other $VP$ decay branching ratios as a maximum rate
are presented. The exclusive results for $t$ and $s$-channel meson
loops and SOZI processes are also listed.  We also include
isospin-violating channels $J/\psi\pi^0$, $\omega\pi^0$,
$\rho^0\eta$, and $\rho^0\eta^\prime$, which can be recognized via
the non-exact cancelations between the charged and neutral meson
loop amplitudes due to the mass differences between the charged and
neutral intermediate mesons.  We do not consider $\phi\pi^0$ channel
since it involves both OZI doubly disconnected process and isospin
violation, thus will be strongly suppressed.

\begin{table}[ht]
\caption{Branching ratios for $\psi(3770)\to VP$ calculated for
different mechanisms. The values for $J/\psi\eta$ and $\phi\eta$ are
fixed at the central values of the experimental
data~\protect\cite{pdg2008}, and the experimental upper limit is
taken for $\rho\pi$~\cite{Ablikim:2005cd}. }\label{tab-1}
\begin{tabular}{|c|c|c|c|c|c|}
\hline BR$(\times 10^{-4})$  &$t$-channel  & $s$-channel & SOZI & Total    \\[1ex]
\hline  $J/\psi\eta$           & $8.44$                & $0.13$                      & --          & $9.0$                \\[1ex]
\hline  $J/\psi\pi^0$          & $0.1 $                & $2.58\times
10^{-2}$ & -- & $4.4\times 10^{-2}$
\\[1ex]
\hline  $\rho\pi$              & $34.45$               & $7.69\times 10^{-5}$       & $8.53$                & $24.0$  \\[1ex]
\hline  $K^{*+} K^- + c.c$     & $10.97$               & $6.83\times 10^{-6}$       & $5.72$                & $8.91$  \\[1ex]
\hline  $K^{*0} {\bar K}^0+c.c$& $11.80$               & $4.38\times 10^{-5}$       & $5.72$                & $9.90$  \\[1ex]
\hline  $\phi\eta$             & $1.25$                & $1.13\times 10^{-5}$       & $1.16$                & $3.1$   \\[1ex]
\hline  $\phi\eta^\prime$      & $0.87$                & $2.53\times 10^{-5}$       & $1.86$                & $3.78$  \\[1ex]
\hline  $\omega\eta$           & $6.83$                & $9.64\times 10^{-6}$       & $1.88$                & $4.69$  \\[1ex]
\hline  $\omega\eta^\prime$    & $0.58$                & $2.87\times 10^{-5}$       & $0.97$                & $0.39$   \\[1ex]
\hline  $\rho\eta$             & $1.88\times 10^{-2}$  & $1.77\times 10^{-5}$       & --           & $1.8\times 10^{-2}$  \\[1ex]
\hline  $\rho\eta^\prime$      & $1.08\times 10^{-2}$  & $1.54\times 10^{-5}$       & --           & $1.0\times 10^{-2}$  \\[1ex]
\hline  $\omega\pi^0$          & $2.57\times 10^{-2}$  & $1.82\times 10^{-5}$       & --           & $2.5\times 10^{-2}$  \\[1ex]
\hline  Sum          & $75.34$  & $0.16$  & $25.84$ &
$63.87$ \\
\hline
\end{tabular}
\end{table}

The following points can be learned from Tab.~\ref{tab-1}: (i)
Different from the $\psi(2S)$-$\psi(1D)$ mixing scheme discussed in
Refs.~\cite{Ding:1991vu,Rosner:2001nm}, our $s$-channel
$\psi(3770)\to \psi^\prime$ transition element is a complex number.
If we neglect the imaginary part due to the widths, we can extract
the mixing angle $\phi\simeq 4.57^\circ$ in the convention of
\cite{Rosner:2001nm}. We find that the $t$-channel transitions are
much more important in $\psi(3770)\to VP$, while the $s$-channel
contributions are generally small and even negligible in light $VP$
channels. This is mainly due to the small partial widths for
$\psi^\prime$ decays into light $VP$. The only non-negligible
$s$-channel is in $\psi(3770)\to J/\psi\eta$, which adds to the
$t$-channel constructively. In contrast, the isospin violating
channel $J/\psi\pi^0$ experiences a destructive interference between
the $t$ and $s$-channel. These results are useful for clarifying the
scenario of $\psi(2S)$-$\psi(1D)$ mixing. (ii) The SOZI coupling
$g_S$ and phase angle $\delta$ are strongly correlated. Applying the
BES data~\cite{Ablikim:2005cd}, we find that the meson loop and SOZI
amplitudes have constructive interferences in $\phi\eta$ and
$\phi\eta^\prime$, but have destructive interferences in $\rho\pi$,
$K^*\bar{K}+c.c.$, and $\omega\eta(\eta^\prime)$, which are
automatically given by the SU(3) flavor symmetry. This is a strong
constraint for our model parameters, and a sum over the $VP$ decays
gives a rate of $\sim 0.64\%$. By varying $\delta$, but keeping the
$\phi\eta$ rate unchanged (i.e. $g_S$ will be changed), we obtain a
lower bound for the sum of branching ratios, $\sim 0.41\%$.

It is interesting to see that the intermediate $D$ meson
rescatterings indeed account for some deficit for the non-$D\bar{D}$
decay. In order to clarify this puzzling problem, it is essential to
have precise data for $\rho\pi$ and $K^*\bar{K}+c.c.$ A search for
these decays at BES-III~\cite{Asner:2008nq} is thus strongly
recommended. Theoretical investigation of other channels such as
$\psi(3770)\to VS$, $VT$, etc is also needed as a prediction and
test of the proposed mechanism.


We thank B. Heltsley, H. Muramatsu, and C.Z. Yuan for useful
communications on CLEO and BES results. This work is supported, in
part, by the National Natural Science Foundation of China (Grants
No. 10675131 and 10491306), Chinese Academy of Sciences
(KJCX3-SYW-N2), and the U.K. EPSRC (Grant No. GR/S99433/01).

{\it Notes added:} \ We would also like to mention that upon the
submission of this paper, a work based on a similar idea was
submitted to the arXiv by Liu {\it et al}~\cite{Liu:2009dr}. There,
the authors focus on the intermediate $D\bar{D}$ rescattering in an
on-shell approximation and investigate its contributions to
$J/\psi\eta$, $\rho\pi$ and $J/\psi\pi\pi$. In our case, we
calculate all $VP$ channels with full loop integrals and a
reasonable estimate of the SOZI processes based on a stringent
constraint on the model parameters.

\end{document}